\documentclass[fleqn]{2023SCGE}
\usepackage[version=4]{mhchem}

\begin{document}

\ensubject{subject}

\ArticleType{Article}
\SpecialTopic{SPECIAL TOPIC: }
\Year{2023}
\Month{January}
\Vol{66}
\No{1}
\DOI{??}
\ArtNo{000000}
\ReceiveDate{January 11, 2023}
\AcceptDate{April 6, 2023}

\title{Magnetism of \ce{NaYbS2}: From finite temperatures to ground state}{Magnetism of \ce{NaYbS2}: From finite temperatures to ground state}

\author[1,2]{Weizhen\,Zhuo}{}
\author[2]{Zheng\,Zhang}{{zhangzheng@iphy.ac.cn}}
\author[1,2]{Mingtai\,Xie}{}
\author[1]{\\Anmin\,Zhang}{}
\author[2]{Jianting\,Ji}{}
\author[2]{Feng\,Jin}{}
\author[1,2]{Qingming\,Zhang}{}

\AuthorMark{Zheng\,Zhang}

\AuthorCitation{Zhuo W Z, Zhang Z, Xie M T, et al}

\address[1]{School of Physical Science and Technology, Lanzhou University, Lanzhou 730000, China}
\address[2]{Beijing National Laboratory for Condensed Matter Physics, Institute of Physics, Chinese Academy of Sciences, Beijing 100190, China}

\abstract{Rare-earth chalcogenide compounds \ce{ARECh2} (A = alkali or monovalent metal, RE = rare earth, Ch = O, S, Se, Te) are a large family of quantum spin liquid (QSL) candidate materials. \ce{NaYbS2} is a representative member of the family. Several key issues on \ce{NaYbS2}, particularly how to determine the highly anisotropic spin Hamiltonian and describe the magnetism at finite temperatures and the ground state, remain to be addressed. In this paper, we conducted an in-depth and comprehensive study on the magnetism of \ce{NaYbS2} from finite temperatures to the ground state. Firstly, we successfully detected three crystalline electric field (CEF) excitation energy levels using low-temperature Raman scattering technique. Combining them with the CEF theory and magnetization data, we worked out the CEF parameters, CEF energy levels, and CEF wavefunctions. We further determined a characteristic temperature of $\sim$40 K, above which the magnetism is dominated by CEF excitations while below which the spin-exchange interactions play a main role. The characteristic temperature has been confirmed by the temperature-dependent electron spin resonance (ESR) linewidth. Low-temperature ESR experiments on the dilute magnetic doped crystal of \ce{NaYb_{0.1}Lu_{0.9}S2} further helped us to determine the accurate $g$-factor. Next, we quantitatively obtained the spin-exchange interactions in the spin Hamiltonian by consistently simulating the magnetization and specific heat data. Finally, the above studies allow us to explore the ground state magnetism of \ce{NaYbS2} by using the density matrix renormalization group. We combined numerical calculations and experimental results to demonstrate that the ground state of \ce{NaYbS2} is a Dirac-like QSL.}



\maketitle

\begin{multicols}{2}
\section{Introduction}\label{section1}
The quantum spin liquid (QSL) is a novel state that the spins exhibit a high degree of entanglement. Unlike conventional magnets, these entangled spins do not form long-range magnetic order even down to zero temperature. This behavior arises due to the influence of strong quantum fluctuations\cite{Balents2010,Savary_2017,ANDERSON1973153,La2CuO4}. 
Initial research efforts on QSL candidate materials primarily centered on transition metal compounds.\Authorfootnote

\noindent
For example, the copper-based magnetic material \ce{ZnCu3(OH)6Cl2}\cite{PhysRevLett.98.077204,PhysRevLett.100.087202,PhysRevB.76.132411,PhysRevLett.101.026405} with a Kagome lattice was the first to be extensively studied as a QSL candidate material.
Inelastic neutron scattering (INS) experiments show that its ground state has a continuous excitation spectrum related to QSL\cite{Han2012}.
Organic materials forming triangular lattice are also promising QSL candidates, such as \ce{\kappa-(BEDT-TTF)2Cu2(CN)3}\cite{doi:10.1126/science.abc6363,Yamashita2009,Isono2016} and \ce{EtMe3Sb[Pd(dmit)2]2}\cite{PhysRevLett.123.247204,PhysRevB.84.094405,PhysRevB.101.140407}.
Thermal conductivity experiments show that \ce{\kappa-(BEDT-TTF)2Cu2(CN)3} has non-zero thermal conductivity even as the temperature approaches 0 K. This observation suggests the presence of itinerant spinon excitations within the material.

In recent years, rare earth materials with triangular lattice have attracted great attention.
Firstly, the triangular lattice inherently possesses a geometric frustration effect, which provides a key structure foundation for realizing QSL.
Secondly, rare earth ions have strong spin-orbit coupling (SOC). 
Based on the Jackeli-Khaliullin scenario\cite{PhysRevLett.101.216804,PhysRevLett.105.027204}, anisotropic spin-exchange interactions can be produced through strong SOC.
More importantly, rare earth ions with an odd number of 4f electrons are protected by time reversal symmetry, namely Kramers doublets.
The Kramers doublets are robust and less affected by lattice distortions or other structural defects.
\ce{YbMgGaO4}\cite{Li2015,PhysRevLett.118.107202,PhysRevLett.115.167203,PhysRevX.8.031028,Li2017,Paddison2017,PhysRevLett.122.137201} is the first discovered triangular rare earth frustrated material satisfying the aforementioned requirements.
Experimental data indicates that \ce{YbMgGaO4} exhibits strong magnetic anisotropy. 
Crucially, the energy gap between the crystalline electric field (CEF) ground state and the first excited state in \ce{YbMgGaO4} exceeds 30 meV\cite{PhysRevLett.118.107202}, allowing the doubly degenerate CEF ground state to be mapped to an effective spin-1/2 model.
A large amount of experimental evidence, including INS\cite{Paddison2017,PhysRevLett.122.137201}, muon spin relaxation ($\mu$SR)\cite{PhysRevLett.117.097201}, and thermodynamic measurements, consistently support the ground state of \ce{YbMgGaO4} as a gapless QSL state.
Therefore, \ce{YbMgGaO4} is considered as a typical QSL material.
However, the previously discussed materials all exhibit limitations to some extent.
For example, there are serious problems of ion vacancies and substitutions in transition metal magnetic materials, which is related to the valence change of transition metals.
In \ce{YbMgGaO4}, there is a problem of \ce{Mg} and \ce{Ga} disorder, which can cause charge randomness and possibly affect the ground state magnetism\cite{https://doi.org/10.1002/qute.201900089}.

In recent years, a large family of QSL candidate rare-earth chalcogenide compounds \ce{ARECh2} (A = alkali or monovalent metal, RE = rare earth, Ch = O, S, Se, Te) has been discovered\cite{Liu_2018, PhysRevB.99.180401, Bordelon2019,PhysRevB.100.144432,PhysRevB.98.220409,PhysRevB.100.224417,PhysRevB.103.035144,PhysRevB.103.184419,PhysRevX.11.021044,PhysRevB.106.085115,Wu2022}.
Within this family, the sub-family \ce{NaYbCh2} (Ch = O, S, Se, Te) is particularly noteworthy, and has the following advantages.
Firstly, compared to \ce{YbMgGaO4}, the sub-family eliminates the Mg/Ga disorder while maintaining the perfect triangular lattice. 
Secondly, through solid-phase reactions, high-quality polycrystalline or single-crystal samples can be grown\cite{Liu_2018}.
Thirdly, this sub-family of materials has excellent two-dimensional properties, and can be achieved with few or single layers through simple mechanical exfoliation.
Fourth, the regulation of magnetic ions or coordination ions can be easily achieved by doping or replacement, which provides a good platform for in-depth study of the evolution mechanism of QSL.
\ce{NaYbO2} is a typical member of this sub-family. A series of experimental results, including magnetization, thermodynamic measurements, INS, and $\mu$SR experiments, all point to a spin disordered ground state for \ce{NaYbO2}\cite{Bordelon2019,PhysRevB.100.144432}.
However, the lack of single crystal samples of this material hinders a deeper study and understanding of its ground state magnetism.
\ce{NaYbSe2} is the most extensively studied in the sub-family, an important reason being that centimeter-size single crystals can be obtained through solid-phase reactions.
The INS experiments observed features of a spinon Fermi surface in \ce{NaYbSe2}\cite{PhysRevX.11.021044}, which has attracted widespread attention in the field of frustrated magnetism.
Adding to its intrigue, the energy gap from the valence band to the conduction band of \ce{NaYbSe2} is about 1.9 eV\cite{Liu_2018}, and metallization and even superconductivity can be achieved by applying high pressure\cite{zhang2020pressure,Jia2020}.
This has advanced one step further in revealing the electronic pairing mechanism of QSL and high-temperature superconductivity.
There are also many experimental reports on \ce{NaYbS2}\cite{PhysRevB.98.220409,PhysRevB.100.241116,Sichelschmidt_2019,Wu2022}. However, some key information about \ce{NaYbS2} is not fully clear yet.
Firstly, three CEF energy levels of \ce{NaYbS2} have not been fully determined\cite{PhysRevB.98.220409}. A quantitative description of the CEF is lacking in \ce{NaYbS2}.
Secondly, despite some experimental investigation into the $g$-factor of \ce{NaYbS2}\cite{Sichelschmidt_2019, PhysRevMaterials.6.046201}, further refinement remains necessary. This is particularly crucial considering the potential influence of the effective field arising from spin-exchange interactions on the $g$-factor.
Another more critical piece of information is the lack of determination of the spin-exchange interactions in the low-energy Hamiltonian.
These interactions are very important for studying the ground state magnetism of \ce{NaYbS2}.
There is also a lack of corresponding numerical calculations for the magnetism of the ground state of \ce{NaYbS2}.
Therefore, \ce{NaYbS2} deserves further and comprehensive study.

In this paper, we conducted an in-depth study on the magnetism of \ce{NaYbS2} from finite temperatures to the ground state. Firstly, we detected the two optical phonon peaks and three CEF excitation energy levels of \ce{NaYbS2} based on polarized Raman spectroscopy and low-temperature Raman scattering. For the three CEF excitations, they are at 207 cm$^{-1}$ ($\sim$ 25.6 meV), 263 cm$^{-1}$ ($\sim$ 32.6 meV), and 314 cm$^{-1}$ ($\sim$ 38.9 meV), respectively.
Secondly, we obtained the CEF parameters, excitation energy levels, and wave functions by fitting the temperature-dependent magnetization ($M/H-T$) data of \ce{NaYbS2} under a magnetic field of 0.1 T. In the fitting process, we also confirmed that the characteristic temperature of \ce{NaYbS2} is 40 K.
Above the characteristic temperature, the CEF excitations play important roles in magnetism. Below the temperature, the spin-exchange interaction dominates.  
ESR measurements on the dilute doped sample of \ce{NaYb_{0.1}Lu_{0.9}S2} yielded precise $g$-factor values of 3.14 in the ab-plane and 0.86 along the c-axis.
We also explained the influence of the effective field contributed by the spin-exchange interaction on the $g$-factor based on the mean-field (MF) theory.
Next, by fitting $M/H-T$ data under a magnetic field of 14 T below the characteristic temperature, we obtained the diagonal spin-exchange interactions $J_{\pm}$ and $J_{zz}$ to be 3.52 K and $-$0.87 K, respectively. The off-diagonal spin-exchange interactions $J_{\pm\pm}$ and $J_{z\pm}$ were determined to be $-$0.18 K and 2.66 K, respectively, by fitting the specific heat data in the low temperature range.
The most important thing is that we studied the ground state magnetism of \ce{NaYbS2}.
We built the ground state phase diagram using the density matrix renormlization group (DMRG) method. We found that spin-exchange interactions of \ce{NaYbS2} are located in the QSL region of the phase diagram and the $S\left( \vec{q} \right)$ in reciprocal space exhibits a broadened peak at the high-symmetry K point.
These calculations preliminarily reveal that the ground state of \ce{NaYbS2} is a Dirac-like QSL state, consistent with the neutron reflections and INS experiment. It also lays the foundation for further studying the potential topological excitations of the \ce{NaYbS2} ground state.

\section{Materials and Method}
The single-crystal \ce{NaYbS2}, \ce{NaLuS2} and \ce{NaYb_{0.1}Lu_{0.9}S2} were prepared using the NaCl-flux method\cite{Liu_2018,PhysRevB.103.035144}.
The typical size of a single crystal is 5 $\times$ 5 $\times$ 0.05 mm$^{3}$.
The high quality of these single-crystal samples was confirmed by X-ray diffraction (XRD) and energy dispersive x-ray spectroscopy (EDX) \cite{Liu_2018}(see Supporting Information S1 and S2).

$\sim$ 1.75 mg of \ce{NaYbS2} single crystals were prepared for M/H-T and magnetic field-dependent magnetization (M-H) measurements.
The anisotropic measurements along the c-axis and in the ab-plane were performed using a Quantum Design DynaCool Physical Property Measurement System (PPMS) with vibration sample meter (VSM) from 1.8 to 300 K under magnetic fields ranging from 0 T to 14 T. 

The zero-field heat capacity data of single crystal sample \ce{NaLuS2} ($\sim$ 0.93 mg) were performed using PPMS at 1.8 $\sim$ 100 K.

A single crystal of \ce{NaYbS2} with size of 5 $\times$ 5 $\times$ 0.05 mm$^{3}$ was used for polarized Raman spectra measurements and low-temperature Raman scattering at 2 K.
These data were collected with a Jobin Yvon LabRAM HR Evolution spectrometer equipped with a liquid-nitrogen-cooled back-illuminated charge-coupled device detector and a liquid-helium-free cryostat. About 1 mW of laser power at 473 nm was focused into a spot with a diameter of $\sim$ 5 $\mu$m on the sample surface. The single crystal of \ce{NaYbS2} was mechanically exfoliated to obtain a clean surface before measurement.

ESR measurements were performed on the single crystals of \ce{NaYbS2} and \ce{NaYb_{0.1}Lu_{0.9}S2} with a Bruker EMX plus 10/12 continuous-wave spectrometer at X-band frequencies (${f \sim }$ 9.39 GHz).
These measurements were carried out both in the ab-plane and along the c-axis from 1.8 K to 70 K.
The angle rotation ESR experiments were also conducted on the spectrometer from 0$^{\circ}$ (parallel to the ab-plane) to 90$^{\circ}$ (parallel to the c-axis).

We formulated MF approximation for anisotropic spin Hamiltonian and derived self-consistent equations to simulate $M/H-T$ data.  3 $\times$ 4 spin sites FD with periodic boundary conditions (PBC) were employed for the specific heat simulation. 
We applied the DMRG method\cite{itensor,RevModPhys.77.259} to calculate the ground state phase diagram of anisotropic spin Hamiltonian and ground state magnetism of \ce{NaYbS2}.
The PBC was employed to get accurate DMRG results. The cluster used in the DMRG simulations is $L_{x}$ $\times$ $L_{y}$ = 12 $\times$ 12 and $L_{x}$ $\times$ $L_{y}$ = 15 $\times$ 15. The truncation error was kept $\sim$ 10$^{-5}$ and performed 20 sweeps to improve the accuracy of the simulations.
The largest bond dimension is $M$ $\sim$ 45000 in the DMRG calculation. 

\begin{figure}[H]
	\centering
	\includegraphics[scale=0.45]{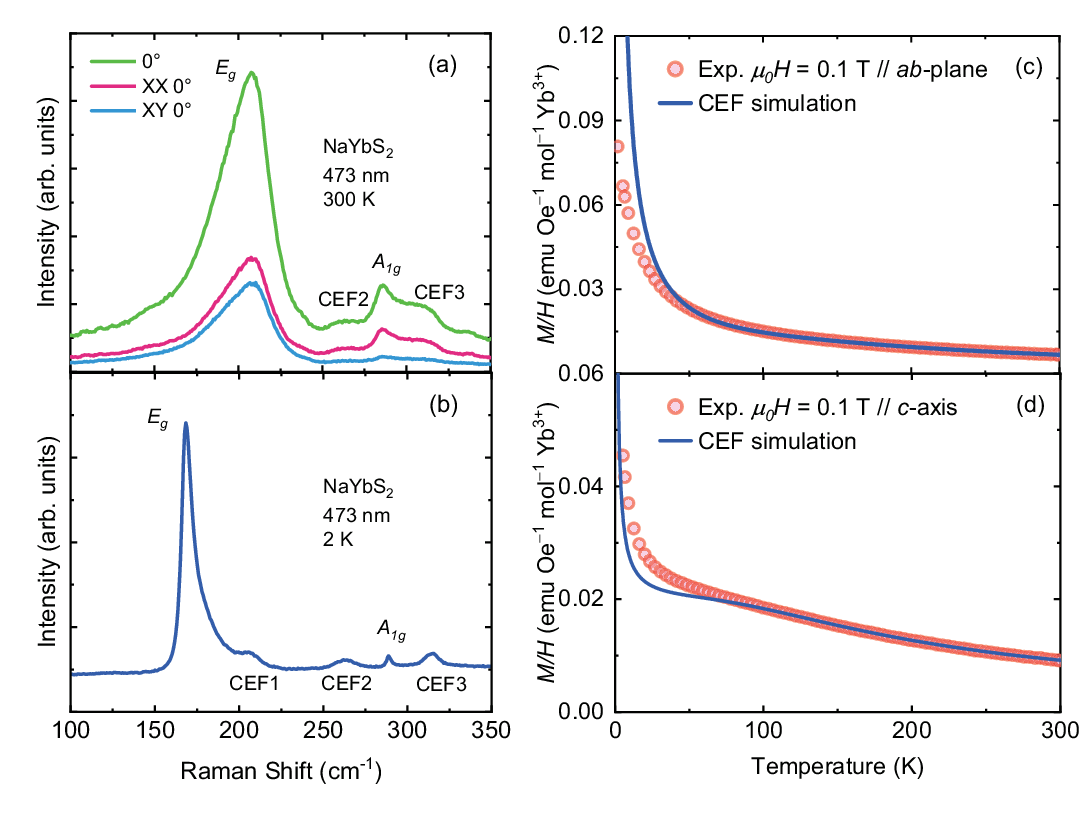}
	\caption{Raman scattering spectra and $M/H-T$ data for \ce{NaYbS2}. (a) The polarized Raman scattering spectra of \ce{NaYbS2} at 300 K. XX and XY represent parallel and cross polarization configurations, respectively. (b) The Raman scattering spectra of \ce{NaYbS2} at 2 K. The $M/H-T$ curves in the ab-plane (c) and along the c-axis (d) under a magnetic field of 0.1 T. The red open circles and blue solid lines are experimental data and simulation results based on the CEF theory, respectively.}
	\label{fig1}
\end{figure}

\section{Effective magnetic Hamiltonian}
The electron configuration of 4$f$ orbital in \ce{Yb^{3+}} is $4f^{13}$. 
Following the Hund's rules, there emerges a spectral term $^{2}I_{7/2}$ featuring an 8-fold degeneracy and another spectral term $^{2}I_{5/2}$ with a 10-fold degeneracy after SOC.
The energy gap between the two spectral terms is approximately 1 eV \cite{PhysRevB.100.174436}, suggesting that transitions between SOC energy levels can be neglected even at room temperatures. 
Consequently, the spectral term $^{2}I_{7/2}$ denotes the SOC ground state configuration.
Considering the shielding effect of the $6s$ and $5p$ orbitals on the $4f$ electrons, the CEF environment formed by surrounding $S^{2-}$ ligands with $D_{3d}$ point group symmetry plays a crucial role in \ce{NaYbS2}.
In this CEF environment, \ce{Yb^{3+}} ions with an odd number of $4f$ electrons split into four pairs of doubly degenerate Kramers states, protected by time-reversal symmetry.
The INS experiments on \ce{NaYbS2} observed the first CEF excitation at approximately 23 meV ($\sim$ 270 K)\cite{PhysRevB.98.220409}. 
This implies that excitation effects of the CEF cannot be ignored when the temperature is comparable to the excitation energy.
In the limit of temperatures significantly lower than the first CEF energy level, the influence of excitations becomes negligible. 
This allows for the construction of an effective spin-1/2 model capturing the essential physics. Within this model, spin-exchange interactions, correlations, and quantum fluctuations take center stage.

Based on the understanding of the \ce{Yb^{3+}} in \ce{NaYbS2} described above, we can construct a magnetic effective Hamiltonian to describe the magnetism of \ce{NaYbS2}.
\ce{NaYbS2} and \ce{NaYbSe2}\cite{PhysRevB.103.035144,PhysRevB.103.184419} share the same triangular lattice formed by \ce{Yb^{3+}} ions due to their isostructural crystal structure.
Therefore, the magnetic effective Hamiltonian used to describe the magnetism of \ce{NaYbS2} within the temperature range we studied is represented as follows\cite{PhysRevB.94.035107,PhysRevB.103.184419}: 
\begin{align}
	\hat{H}_{\mathrm{eff}}= & \hat{H}_{\mathrm{CEF}}+\hat{H}_{\mathrm{spin}-\text { spin }}+\hat{H}_{\text {Zeeman }} \notag \\
	= & \sum_i \sum_{m, n} B_m^n \hat{O}_m^n \notag\\
	& +\sum_{\langle i j\rangle}\left[J_{z z} S_i^z S_j^z+J_{ \pm}\left(S_i^{+} S_j^{-}+S_i^{-} S_j^{+}\right)\right. \notag\\
	& +J_{ \pm \pm}\left(\gamma_{i j} S_i^{+} S_j^{+}+\gamma_{i j}^* S_i^{-} S_j^{-}\right) \notag\\
	& \left.-\frac{i J_{z \pm}}{2}\left(\gamma_{i j} S_i^{+} S_j^z-\gamma_{i j}^* S_i^{-} S_j^z+\langle i \longleftrightarrow j\rangle\right)\right] \notag\\
	& -\mu_0 \mu_B \sum_i\left[g_{a b}\left(h_x S_i^x+h_y S_i^y\right)+g_c h_c S_i^z\right], \label{1eq}
\end{align}
where $B_m^n$ denotes the CEF parameters and $O_{m}^{n}$ symbolizes the Steven operator, which is constructed based on the angular momentum $\hat{J}$ after the SOC. $S_i^\alpha (\alpha = x,y,z)$ is the effective spin-1/2 operators at site $i$ and $S_i^{\pm} = S_i^x \pm iS_i^y$ are the non-Hermitian ladder operators. The nearest neighbor (NN) anisotropic spin-exchange interactions are denoted by $J_{\pm}$, $J_{zz}$, $J_{\pm\pm}$, and $J_{z\pm}$. The phase factor $\gamma_{ij}$ is taken as 1, $\mathrm{e}^{i \frac{2\pi}{3}}$ and $\mathrm{e}^{-i \frac{2\pi}{3}}$ along the three bonds $\vec{a}_1, \vec{a}_2$ and $\vec{a}_3$. The Land$\text{\'{e}}$ $g$-factors in the ab-plane and along the c-axis are represented by $g_{ab}$ and $g_c$, respectively.

It is noteworthy that the CEF part and the anisotropic spin-exchange interactions part in the magnetic effective Hamiltonian do not exist in the same Hilbert space. 
The CEF description is particularly effective for capturing the high-temperature magnetism dominated by CEF excitations. Conversely, the spin-exchange interaction model is suitable for describing low-energy spin excitations within the CEF ground state.
Therefore, the analysis of magnetization and thermodynamic data needs to be discussed separately in two parts.
In the following, we would apply the above strategy to study the magnetism of \ce{NaYbS2} from finite temperatures to the ground state.

\section{CEF and Raman spectra}
For the study of CEF excitations in rare-earth materials, INS offers a powerful capability. This technique can directly probe CEF energy levels that are independent of momentum transfer $\vec{Q}$.
Besides, Raman scattering provides another valuable tool for investigating CEF excitations. Compared to INS, Raman scattering offers the advantage of more convenient laboratory-based measurements of CEF excitations.
In \cref{fig1}(b), we show the Raman spectrum of \ce{NaYbS2} at 2 K.
From the spectrum, we can clearly observe five excitation peaks in the range of 100 to 350 cm$^{-1}$.
Among these peaks, we can divide them into two parts: one part corresponds to the phonon peaks related to optical branch lattice vibrations, and the other part corresponds to the CEF peaks associated with CEF excitations.
By employing both symmetry analysis and polarization scattering experiments, we can determine the vibrational modes associated with the observed phonon peaks.
With the spatial symmetry of $R\overline{3}m$, Raman scattering can detect two phonon modes in \ce{NaYbS2}, namely $A_{g}$ and $E_{g}$ modes.
The Raman scattering tensors of $A_{g}$ and $E_{g}$ modes tell us that the $A_{g}$ mode is visible only in the parallel polarization (XX) configuration, while the $E_{g}$ mode can be observed in both XX and cross polarization (XY) configurations.
The polarized Raman spectrum of \ce{NaYbS2} at 300 K are shown in \cref{fig1}(a). The two modes can be clearly identified and consistent with our symmetry analysis.

\begin{table*}
	\caption{\label{tab:table2}The fitting results of CEF ${B^{n}_{m}}$ parameters, energy and wave functions for $\mathrm{NaYbS_2}$}

		\begin{tabular}{lcccccc}
			\hline
			&${B^{0}_{2}}$ &  ${ B^{0}_{4}}$ & ${B^{3}_{4}}$ &  ${B^{0}_{6}}$ & ${B^{3}_{6}}$ & ${B^{6}_{6}}$
			\\ \hline
			${B^{n}_{m}}$ (meV)&${-1.409\times10^{-1}}$& ${2.302\times10^{-3}}$ &${-2.175\times10^{-1}}$ & ${6.478\times10^{-4}}$ & ${-5.230\times10^{-3}}$ & ${1.765\times10^{-2}}$ 
			
			\\ \hline
			{Energy (meV) } & \multicolumn{5}{c} {Wave function} \\ \hline
			
			0.0 & $-0.639(1)\left|-\frac{7}{2}\right\rangle$ & $+0.113(6)\left|-\frac{5}{2}\right\rangle$ & $-0.353(9)\left|-\frac{1}{2}\right\rangle$ & $+0.060(8)\left|\frac{1}{2}\right\rangle$ &
			$+0.661(5)\left|\frac{5}{2}\right\rangle$ &
			$-0.109(8)\left|\frac{7}{2}\right\rangle$ \\
			0.0 & $-0.109(8)\left|-\frac{7}{2}\right\rangle$ & $-0.661(5)\left|-\frac{5}{2}\right\rangle$ & $-0.060(8)\left|-\frac{1}{2}\right\rangle$ & $-0.353(9)\left|\frac{1}{2}\right\rangle$ &
			$+0.113(6)\left|\frac{5}{2}\right\rangle$ &
			$+0.639(1)\left|\frac{7}{2}\right\rangle$\\%

			24.449(6) & $-0.038(8)\left|-\frac{7}{2}\right\rangle$ & $-0.366(8)\left|-\frac{5}{2}\right\rangle$ & $-0.598(3)\left|-\frac{1}{2}\right\rangle$ & $+0.613(7)\left|\frac{1}{2}\right\rangle$ &
			$-0.357(5)\left|\frac{5}{2}\right\rangle$ &
			$-0.039(8)\left|\frac{7}{2}\right\rangle$ \\
			
			24.449(6) & $-0.039(8)\left|-\frac{7}{2}\right\rangle$ & $+0.357(5)\left|-\frac{5}{2}\right\rangle$ & $-0.613(7)\left|-\frac{1}{2}\right\rangle$ & $-0.598(3)\left|\frac{1}{2}\right\rangle$ &
			$-0.366(8)\left|\frac{5}{2}\right\rangle$ &
			$+0.038(8)\left|\frac{7}{2}\right\rangle$ \\
			
			32.290(9) & \raisebox{0.85mm}{\rule{1cm}{0.2mm}} &  \raisebox{0.85mm}{\rule{1cm}{0.2mm}} & $+0.185(3)\left|-\frac{3}{2}\right\rangle$ & $-0.982(7)\left|\frac{3}{2}\right\rangle$ & \raisebox{0.85mm}{\rule{1cm}{0.2mm}}
			& \raisebox{0.85mm}{\rule{1cm}{0.2mm}}
			\\
			32.290(9) & \raisebox{0.85mm}{\rule{1cm}{0.2mm}}  &  \raisebox{0.85mm}{\rule{1cm}{0.2mm}} & $-0.982(7)\left|-\frac{3}{2}\right\rangle$ & $-0.185(3)\left|\frac{3}{2}\right\rangle$ & \raisebox{0.85mm}{\rule{1cm}{0.2mm}}
			& \raisebox{0.85mm}{\rule{1cm}{0.2mm}}
			\\

			39.622(7) & $+0.030(2)\left|-\frac{7}{2}\right\rangle$ & $+0.535(4)\left|-\frac{5}{2}\right\rangle$ & $-0.014(7)\left|-\frac{1}{2}\right\rangle$ & $+0.369(1)\left|\frac{1}{2}\right\rangle$ &
			$+0.021(3)\left|\frac{5}{2}\right\rangle$ &
			$+0.758(6)\left|\frac{7}{2}\right\rangle$ \\

			39.622(7) & $-0.758(6)\left|-\frac{7}{2}\right\rangle$ & $+0.021(3)\left|-\frac{5}{2}\right\rangle$ & $+0.369(1)\left|-\frac{1}{2}\right\rangle$ & $+0.014(7)\left|\frac{1}{2}\right\rangle$ &
			$-0.535(4)\left|\frac{5}{2}\right\rangle$ &
			$+0.030(2)\left|\frac{7}{2}\right\rangle$ \\
			
			\hline
		\end{tabular}

\end{table*}

In addition to the two phonon peaks, three additional scattering peaks were detected at 207 $\mathrm{cm^{-1}}$ ($\sim$ 25.6 meV), 263 $\mathrm{cm^{-1}}$($\sim$ 32.6 meV), and 314 $\mathrm{cm^{-1}}$ ($\sim$ 38.9 meV). 
The scattering peaks at 207 cm$^{-1}$ and 314 cm$^{-1}$ are close to the CEF excitations measured by INS around 23 meV and 39 meV\cite{PhysRevB.98.220409}.
The scattering peak at 263 cm$^{-1}$ closely matches another weak peak observed around 32 meV in INS, initially attributed to additional peak caused by impurities in the powder samples\cite{PhysRevB.98.220409}.
After careful analysis, we can confirm that the scattering peak at 263 cm$^{-1}$ is caused by CEF excitation.
First, the single crystal samples prepared for the Raman scattering experiments were obtained using mechanical exfoliation to ensure clear surface. Therefore, we can rule out the influence of impurities in the samples.
Second, the scattering peak near 263 cm$^{-1}$ can still be observed at 300 K despite weakening. The peak position does not shift significantly with temperature, which is consistent with the rule of CEF excitation changing with temperature.
Third, a weak excitation peak, also exhibiting momentum independence, was detected by INS at 32 meV\cite{PhysRevB.98.220409}.
Based on the analysis of our Raman scattering experiments, we can determine that the three CEF excitations levels of \ce{NaYbS2} are 25.6, 32.6, and 38.9 meV. 
In addition, the three CEF energy levels determined by fitting the $M/H-T$ data of \ce{NaYbS2} (see below) are consistent with the CEF excitation energy levels measured by Raman scattering. Cross-validation of these experimental data enhances our confidence in using Raman scattering to determine the three CEF excitation energy levels of \ce{NaYbS2}.

Meanwhile, we found that the phonon peak of $E_{g}$ mode has a strong asymmetry, which is mainly caused by the CEF-phonon coupling. We will conduct a more detailed analysis starting from the microscopic model in our future work.

\begin{figure*}[htb]
	\centering
	\includegraphics[scale=0.85]{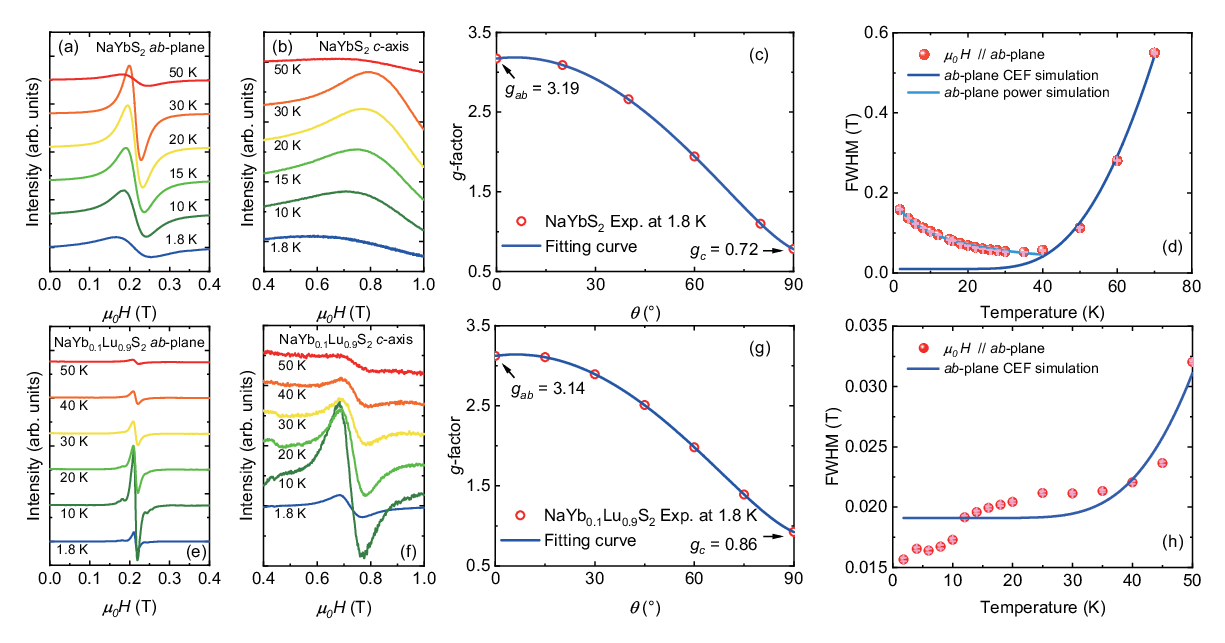}
	\caption{The ESR data of \ce{NaYbS2} and \ce{NaYb_{0.1}Lu_{0.9}S2}. (a) and (b) ESR differential spectra of \ce{NaYbS2} in the ab-plane and along the c-axis. (e) and (f) ESR differential of \ce{NaYb_{0.1}Lu_{0.9}S2} in the ab-plane and along the c-axis. (c) and (g) Angle-dependent $g$-factor of \ce{NaYbS2} and \ce{NaYb_{0.1}Lu_{0.9}S2} at low temperature. (d) and (h) Temperature-dependent ESR linewidth of \ce{NaYbS2} and \ce{NaYb_{0.1}Lu_{0.9}S2} in the ab-plane. The red circles are experimental data and the blue solid lines are simulation results.}
	\label{fig2}
\end{figure*}

\begin{figure*}[htb]
	\centering
	\includegraphics[scale=0.9]{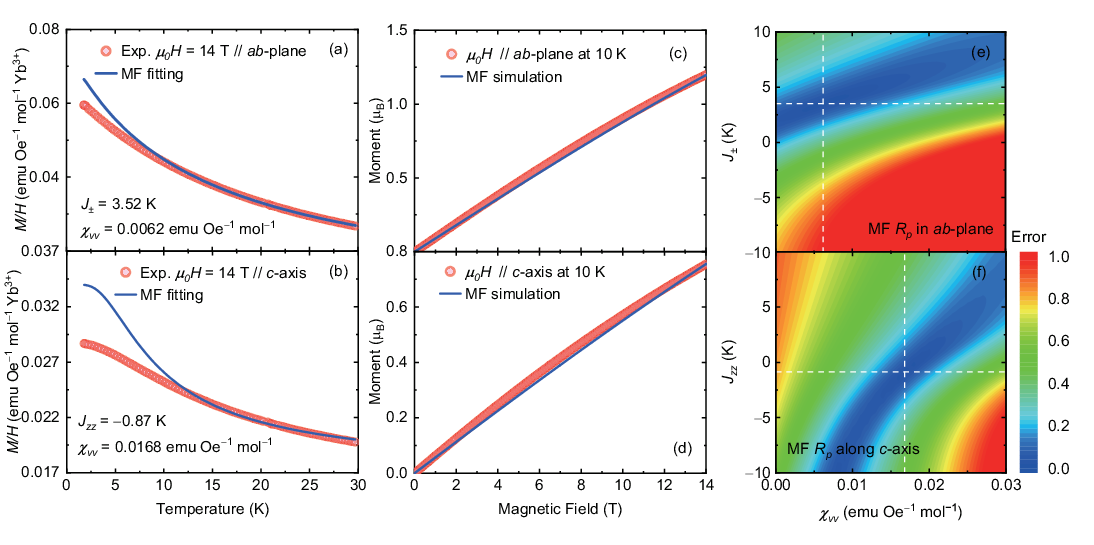}
	\caption{$M/H-T$ and $M-H$ data of \ce{NaYbS2}. The $M/H-T$ data in the ab-plane (a) and along the c-axis (b) under a magnetic field of 14 T. The $M-H$ data in the ab-plane and along the c-axis at 10 K. (e) and (f) Error mapping of $M/H-T$ fitting in the ab-plane and along the c-axis. (c) and (d) share the same color scale in the right corner. Red circles are experimental data and the blue solid lines are simulation results. The intersection marked by the white dashed lines is the optimal position with minimal error.}
	\label{fig3}
\end{figure*}

\section{Magnetism of \ce{NaYbS2} at finite temperatures}
In the previous section, we conducted detailed investigations on the CEF excitations associated with \ce{NaYbS2} based on the low-temperature Raman scattering experiment, accurately determining the three CEF excitation energy levels of \ce{NaYbS2}.
This lays foundations for studying the magnetism of \ce{NaYbS2} at finite temperatures.
In this section, we present a discussion on the magnetism of \ce{NaYbS2} at finite temperatures, including the effect of CEF excitations and the influence of low-energy spin-exchange interactions on magnetism.

We first focus on the contribution of CEF excitations to magnetism.
At high temperature ranges, thermal fluctuations primarily influence magnetization through a series of CEF excitation levels within individual ions.
When the temperature is comparable to the CEF excitation energy levels, the measured magnetization data are mainly contributed by the CEF excitations.
Based on the CEF energy levels obtained from Raman scattering of \ce{NaYbS2} and our measured $M/H-T$ data, we can simulate the $M/H-T$ data by employing magnetization calculation formulas\cite{PhysRevLett.115.167203,PhysRevX.10.011007}.
We used a nonlinear least squares method to fit the contributions of CEF to $M/H-T$ in the ab-plane and along the c-axis. The CEF parameters obtained from the point charge model were used as initial values, and three CEF levels were set as constraints. 
The experimental data of $M/H-T$ and the simulation results based on the CEF theory are shown in \cref{fig1}(c) and \cref{fig1}(d).
A clear discrepancy arises between the simulation results and experimental data below 40 K both in the ab-plane and along the c-axis.
This indicates that the magnetism is mainly contributed by CEF excitations above 40 K. In contrast, the magnetism is predominantly governed by the anisotropic spin-exchange interactions below 40 K.
Therefore, we can define 40 K as the characteristic temperature of \ce{NaYbS2}.
Analyzing the temperature-dependent ESR linewidths can also help us determine the characteristic temperature associated with CEF excitations.
\cref{fig2}(d) and \cref{fig2}(h) show the temperature-dependent of the linewidth for \ce{NaYbS2} and \ce{NaYb_{0.1}Lu_{0.9}S2} samples, respectively. Both figures show that the ESR linewidth rapidly broadens with increasing temperature above 40 K.
This is consistent with the results we obtained by simulating the $M/H-T$ data.
Furthermore, by fitting the ESR linewidth above 40 K with the following formula, we can also estimate the energy gap from the CEF ground state to the first excited state of \ce{NaYbS2}.
\begin{equation}
	\mu_{0}\Delta H \simeq \frac{1}{\mathrm{exp}\left( \Delta E/T \right) - 1}
\end{equation}
where $\Delta H$ is the full width at half maximum (FWHM), $\Delta E$ is the energy gap, and $T$ is the temperature.
The fitting results are shown as the blue solid lines in \cref{fig2}(d) and \cref{fig2}(h), respectively. The $\Delta E$ obtained by fitting estimation is 263 K ($\sim$ 22.67 meV), which is close to the CEF first excitation energy level measured by Raman scattering.
The characteristic temperature of \ce{NaYbS2} is obviously higher than that of \ce{NaYbSe2} (25 K)\cite{PhysRevB.103.184419}.
The primary reason is that \ce{NaYbS2} has higher CEF excitation energy levels compared to \ce{NaYbSe2}, which fundamentally stems from the stronger electronegativity of the \ce{S^{2-}} anion.

In addition, we also obtained the CEF parameters, excitation energy levels, and corresponding wave functions during the non-linear fitting process, as shown in \cref{tab:table2}.
Based on the fitted CEF parameters, the three CEF excitation energy levels obtained through diagonalization calculations are 24.45, 32.29, and 39.62 meV, which are very close to our Raman scattering results. This further demonstrates the reliability of the CEF-related information obtained from our fitting and calculations.

In the limit of temperatures significantly lower than the first CEF excitation energy, the doubly degenerate CEF ground state can be mapped onto an effective spin-1/2 model.
When considering the influence of the magnetic field on the magnetic moment, two additional key parameters arise: the Land$\text{\'{e}}$ $g$-factors in the ab-plane and along the c-axis.
The $g$-factor not only reflects the magnetic anisotropy but also serves as a bridge connecting the CEF ground state and the effective spin-1/2 model.
Therefore, accurately determining the $g$-factors of \ce{NaYbS2} is important.
The methods to determine $g$-factors mainly include two approaches: one method is to obtain the $g$-factors from the saturation magnetization by applying a magnetic field at low temperature, and the other is to directly obtain them from the resonance magnetic field in the ESR spectrum.
Both method have their own advantages and disadvantages.
For the saturation magnetization method, there is no need to consider the influence of effective internal field caused by spin-exchange interactions on the $g$-factors\cite{PhysRevResearch.4.033006}. However, if the interaction between spins is strong or exhibits significant paramagnetism, the magnetic moment may require a very high magnetic field to saturate.
For the ESR method, the $g$-factor can be measured by a weak resonant magnetic field, but it is easily affected by the internal field.
This leads to the $g$-factor measured by ESR to deviate from the true value.
If the $g$-factor of magnetic ions is small, it can also be influenced by spin-exchange interactions, causing the observed ESR linewidth to exceed the range of the spectrometer, and thereby limiting the ability to fully capture the ESR resonance spectrum.
This problem exists in the ESR spectrum along the c-axis of \ce{NaYbS2}, as shown in \cref{fig2}(b).
Therefore, we propose to determine the accurate $g$-factors by measuring the ESR spectrum of a dilute magnetic doped sample.
In \ce{NaYb_{0.1}Lu_{0.9}S2}, spin-exchange interactions are almost nonexistent, thus the $g$-factors obtained from ESR measurements are not affected by spin-exchange interactions.
In \cref{fig2}(e) and \cref{fig2}(f), we respectively present the temperature-dependent ESR resonance spectra of \ce{NaYb_{0.1}Lu_{0.9}S2} in the ab-plane and along the c-axis from 1.8 K to 50 K. 
In dilute magnetic doping samples, issues such as imperfect sample homogeneity and magnetic ion vacancies are inevitably introduced.
The ESR is highly sensitive to these issues, resulting in some subtle features in \cref{fig2}(e).
Based on Lorentzian fitting, we determine the $g$-factors corresponding to the resonance field as $g_{ab} = 3.14$ and $g_{c} = 0.86$.
By substituting the determined $g_{ab}$ and $g_{c}$ into the angle-dependent $g$-factor formula (see Supporting Information S4), we can simulate the experimental measurement results of the $g$-factor at different angles.

To further check the accuracy of our calculations, we also calculated the $g$-factors in the ab-plane and along the c-axis based on the CEF ground state. The expression of the calculation formula is as follows:\cite{PhysRevB.103.035144}:
\begin{align}\label{eq2}
	\begin{gathered}
		g_{ab-\text{plane}}=g_j\left|\left\langle\psi_{ \pm}\left| \hat{J}^{\pm} \right| \psi_{\mp}\right\rangle\right|,  \\
		g_{c-\text{axis}}=2 g_j\left|\left\langle\psi_{ \pm}\left| \hat{J}^{z} \right| \psi_{ \pm}\right\rangle\right|,
	\end{gathered}
\end{align}
where $g_j = 8/7$ and $ |\psi_{ \pm}\rangle$ are the Land$\text{\'{e}}$ $g$-factor of free \ce{Yb^{3+}} and the CEF ground state.
We can obtain the $g_{ab} \sim 3.13$ in the ab-plane and $g_{c} \sim 0.88$ along the c-axis.
This is highly consistent with the results obtained from the ESR spectrum of \ce{NaYb_{0.1}Lu_{0.9}S2}. On the one hand, it demonstrates the reliability of the $g$-factors obtained from the dilute magnetic doped samples. On the other hand, it validates the accuracy of the CEF parameters obtained through fitting the $M/H-T$.
We can compare the $g$-factors of \ce{NaYb_{0.1}Lu_{0.9}S2} with those of \ce{NaYbS2}. We found that the $g$-factor of \ce{NaYb_{0.1}Lu_{0.9}S2} is not significantly different from that of \ce{NaYbS2} ($g_{ab}$ = 3.19) in the ab-plane, but the $g$-factor of \ce{NaYb_{0.1}Lu_{0.9}S2} is significantly larger than that of \ce{NaYbS2} along the c-axis.
The main reasons for this are as follows: In the ab-plane, the spin-exchange interaction $J_{\pm}$ is antiferromagnetic type (see below), and the internal field induced by the antiferromagnetic interaction can counteract each other.

\begin{figure*}[htb]
	\centering
	\includegraphics[scale=0.9]{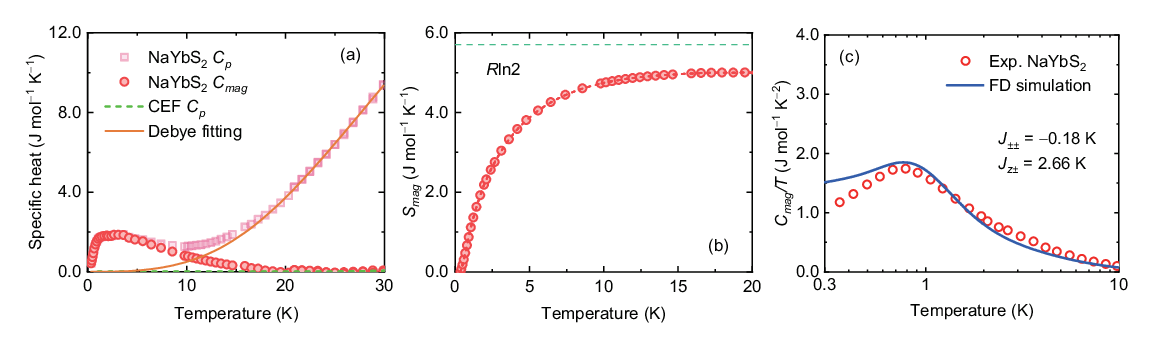}
	\caption{Specific heat of \ce{NaYbS2} (a) Specific heat and magnetic specific heat of \ce{NaYbS2}. The light red squares represent the raw data of the specific heat of \ce{NaYbS2}. The red circles represent the magnetic specific heat of \ce{NaYbS2}. The green dashed line is the calculated CEF specific heat, and the orange solid line represents the results of the lattice vibrational specific heat obtained by fitting the specific heat of \ce{NaYbS2} at temperature range of 20 to 30 K. (b) Magnetic entropy of \ce{NaYbS2}. (c) $C_{mag}/T$ of \ce{NaYbS2} below 10 K and FD simulation results. The red circles are experimental data, and blue solid line is FD simulation result.}
	\label{fig4}
\end{figure*}

The ferromagnetic spin-exchange interaction $J_{zz}$ along the c-axis induces an effective internal field that aligns with the applied magnetic field. Consequently, the superposition of this internal field and the external field leads to a smaller $g_{c}$.
Based on the MF Hamiltonian along the c-axis, we can estimate the magnitude of the internal field.
The MF Hamiltonian along the c-axis is represented as follows:
\begin{equation}
	H_{MF, c-axis} = -g_{c} \mu_{0} \mu_{B} \left(  h_{c} - \frac{6J_{zz} m_{c}}{\mu_{B}^{2} g_{c}^{2}} \right) \sum_{i} S_{i}^{z},
\end{equation}
where $m_{c}$ represents the magnetic moment contributed by the spin-exchange interaction $J_{zz}$.
The relationship between $g_{c}$, resonance frequency, and resonance magnetic field, ignoring spin-exchange interaction $J_{zz}$, can be expressed as follows:
\begin{equation}
	g_{c} = \frac{h\nu}{\mu_{0}\mu_{B} h_{c}},
\end{equation}
where $h$ is planck's constant and $\nu$ is resonance frequency.
Taking into account the effect of spin-exchange interaction $J_{zz}$ on the $g$-factor,
\begin{equation}
	g_{c}' = \frac{h\nu}{\mu_{0} \mu_{B} \left(  h_{c} - \frac{6J_{zz} m_{c}}{\mu_{B}^{2} g_{c}^{2}} \right)}.
\end{equation}
By the ratio of Eq. (5) to (6), we can determine the effective magnetic field of \ce{NaYbS2} contributed by the spin-exchange interaction $J_{zz}$ as
\begin{equation}
	h_{\mathrm{eff}} = \frac{6J_{zz} m_{c}}{\mu_{B}^{2} g_{c}^{2}} = \left( 1 - \frac{g_{c}}{g_{c}'} \right) h_{c} = -\text{0.14 T}
\end{equation}
The negative sign indicates that the exchange interaction $J_{zz}$ along the c-axis is ferromagnetic type, aligning with $M/H-T$ data fitting results at lower temperatures (see below).

So far, we have accurately determined the $g$-factors in the ab-plane and along the c-axis through ESR experiments on \ce{NaYb_{0.1}Lu_{0.9}S2}. We have also explained the reasons for the differences in the $g$-factors between \ce{NaYb_{0.1}Lu_{0.9}S2} and \ce{NaYbS2}.
Next, we can analyze the magnetization and specific heat data below the characteristic temperature based on the MF theory and FD method.
By fitting these data, we can extract the spin-exchange interactions in the spin Hamiltonian, laying the foundation for further investigation into the ground state magnetism of \ce{NaYbS2}.
Due to the limitations of the MF spin Hamiltonian, which is suitable for paramagnetic states, selecting an appropriate magnetic field and a reasonable temperature range is crucial before fitting and simulating the magnetization data.
For \ce{NaYbS2}, magnetic fields can induce a series of phase transitions\cite{Wu2022}. 
From the $M-H$ data of \ce{NaYbS2} at 0.8 K and the magnetic field-temperature ($H-T$) phase diagram, we can see that the magnetic moment gradually saturates as the magnetic field exceeds 12 T. 
Meanwhile, $\mu$SR experiments on \ce{NaYbS2}\cite{PhysRevB.100.241116} reveal that spin fluctuations become insignificant above 8 K.
Therefore, we can confidently select the temperature range of 10 to 30 K under a magnetic field of 14 T to fit the $M/H-T$ data.
In the fitting process, we can extract the diagonal terms of the spin-exchange interactions $J_{\pm}$ and $J_{zz}$ in the MF Hamiltonian.

\begin{figure*}[htb]
	\centering
	\includegraphics[scale=0.87]{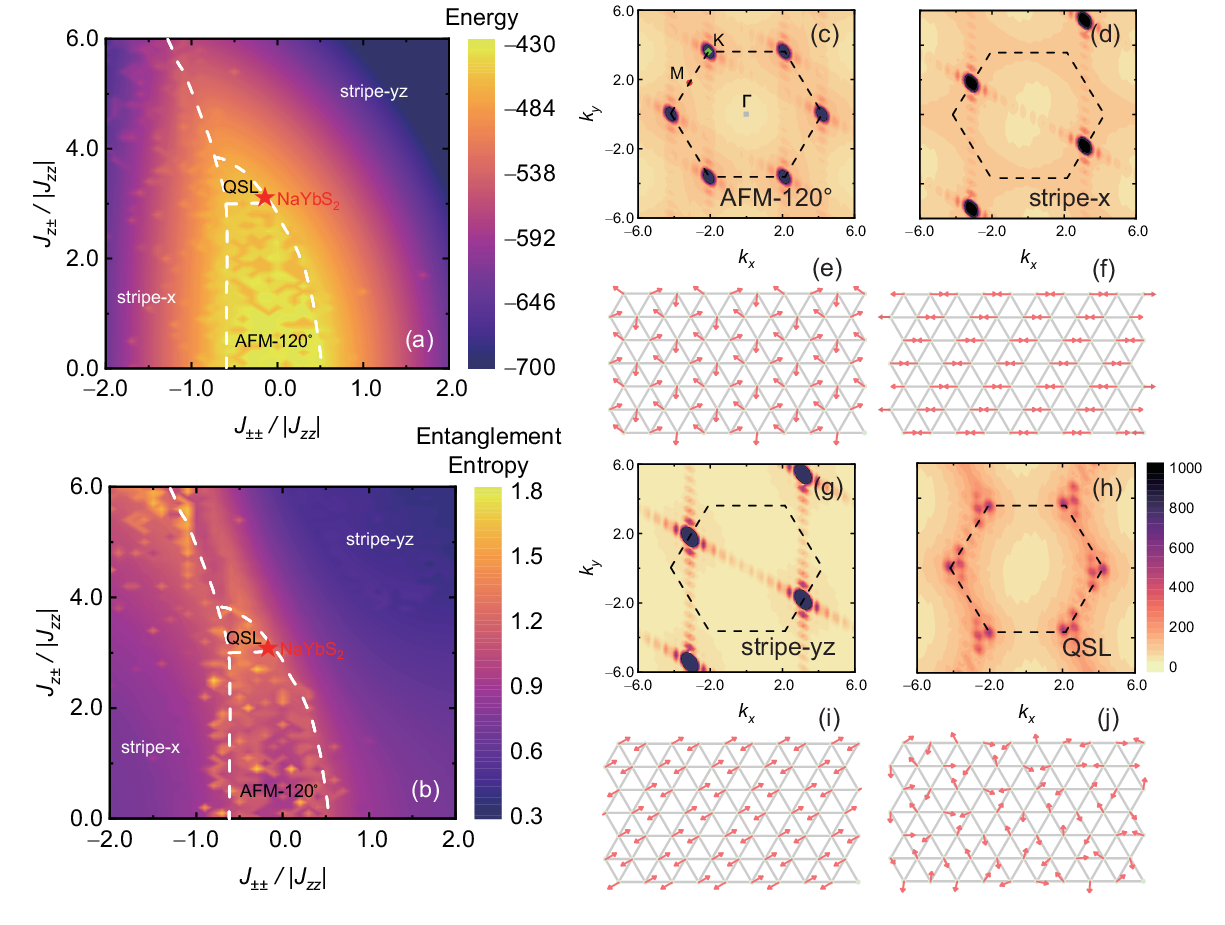}
	\caption{Ground phase diagram, Static spin structure factor $S(\vec{q})$ in reciprocal space, and magnetic structure.  (a) Ground state phase diagram scaled by energy. (b) Ground state phase diagram scaled by entanglement entropy. The white dash lines are boundary of different phases. The red pentagon marks the position of \ce{NaYbS2} in the phase diagram. (c) and (e) $S(\vec{q})$ and magnetic structure of AFM-120$^{\circ}$. (d) and (f)  $S(\vec{q})$ and magnetic structure of stripe-x. (g) and (i)  $S(\vec{q})$ and magnetic structure of stripe-yz. (h) and (j)  $S(\vec{q})$ and magnetic structure of QSL. (c), (d), (g), and (h) share the same color scale in the right.}
	\label{fig5}
\end{figure*}

\cref{fig3}(a) and \cref{fig3}(b) present the $M/H-T$ experimental data and fitting results, including both in the ab-plane and along the c-axis.
From the two figures, we can see that the simulated results are consistent with the experimental measurements within the selected temperature range. However, the simulated results gradually deviate from the experimental data as the temperature drops below 10 K.
The fitting error ($R_{p}$) for the $M/H-T$ data is presented in \cref{fig3}(e) and \cref{fig3}(f).
By searching for the minimum of the fitting error, we can determine $J_{\pm}$, $J_{zz}$, and the Van Vleck (VV) paramagnetism at low temperatures.
In the ab-plane, we obtained $J_{\pm}$ = 3.52 K, $\chi_{VV}$ = 0.0062 emu Oe$\mathrm{^{-1}}$ mol$\mathrm{^{-1}}$, and $R_{p}$ = 0.19 \%.
Along the c-axis, we obtained $J_{zz}$ = -0.87 K, $\chi_{VV}$ = 0.0168 emu Oe$\mathrm{^{-1}}$ mol$\mathrm{^{-1}}$, and $R_{p}$ = 0.40 \%.
We double-checked these spin-exchange interactions by comparing the $M-H$ experimental data with MF simulation results at 10 K.
The $M-H$ data shown in \cref{fig3}(c) and \cref{fig3}(d) are consistent with the simulation results.
From the fitting results, we found that \ce{NaYbS2} exhibits antiferromagnetic-type spin-exchange interaction and a weaker VV paramagnetism in the ab-plane. 
In contrast, it exhibits ferromagnetic-type spin-exchange interaction and a more significant VV paramagnetism along the c-axis.
The paramagnetism also results in nearly linear growth of the magnetic moment along the c-axis under high magnetic fields\cite{PhysRevB.98.220409}.
Besides, spin-exchange interactions can also reflect in the ESR linewidth. As the temperature decreases, the spin-exchange interactions can lead to broadening of the ESR linewidth.
We can use the following empirical formula to fit the ESR linewidth of \ce{NaYbS2} at low temperature range,
\begin{equation}
	\mu_{0} \Delta H \simeq \frac{A}{\left( T - T_{\theta} \right)^p},
\end{equation}
where $A$ is the proportionality coefficient, $p$ is the power coefficient, and $T_{\theta}$ is the Weiss temperature.
The Weiss temperature and spin-exchange interaction $J_{\pm}$ in the ab-plane have the following relationship, which is $T_{\theta} = -3 J_{\pm}$.
By substituting the spin-exchange interaction $J_{\pm}$ into the above formulas to simulate the ESR linewidth in the ab-plane, it can match the experimental data well (see \cref{fig2}(d)). This further indicates that the broadening of the ESR linewidth of \ce{NaYbS2} at low temperatures is caused by the spin-exchange interaction.

For the other two off-diagonal spin-exchange interactions $J_{\pm\pm}$ and $J_{z\pm}$, we can obtain them through fitting the zero-field specific heat data.
We select the specific heat of \ce{NaYbS2} in the temperature range from 0.3 to 10 K\cite{PhysRevB.98.220409}. 
To subtract the contribution of lattice vibrations to the specific heat, we adopted the following strategy.
Considering that \ce{NaYbS2} is an insulator, when the temperature is significantly greater than the spin-exchange interactions ($\ge$ 20 K), the contributions to the specific heat of \ce{NaYbS2} are primarily from two sources: lattice vibrations and CEF excitations. We have determined the CEF parameters, energy levels, and wave functions of \ce{NaYbS2}. Therefore, using thermodynamic formulas, we can directly calculate the contribution of CEF excitations to the specific heat. The calculation results (green dashed line \cref{fig4}(a)) indicate that the contribution of CEF excitations to the specific heat does not exceed 0.1 J mol$^{-1}$ K$^{-1}$ within the temperature range of 20 to 30 K.
Therefore, we can conclude that the specific heat is essentially due to lattice vibrations within this temperature range.
To obtain the contribution of lattice vibration to the specific heat of \ce{NaYbS2}, we employed the Debye formula to fit the specific heat data of \ce{NaYbS2} at the temperature range of 20 to 30 K.
The fitting results are shown by the orange solid line in \cref{fig4}(a).
After subtracting the contribution of lattice vibrations to the specific heat, we obtain the magnetic specific heat of \ce{NaYbS2}, as indicated by the red circles in \cref{fig4}(a).
By integrating the magnetic specific heat of \ce{NaYbS2} over temperature, we can obtain the magnetic entropy of \ce{NaYbS2}, as shown in \cref{fig4}(b).
The figure reveals that the magnetic entropy remains nearly constant above 15 K, yet it has not fully reached $R$ln2, primarily due to the absence of data for lower temperatures ($\le$ 0.35 K).
We also use the Debye formula to fit the specific heat data of \ce{NaLuS2} from 1.8 to 20 K (see Supporting Information S7).
Next, we can extract off-diagonal spin interactions from the magnetic specific heat data by selecting appropriate numerical calculation method.
We employ the FD to fit the magnetic specific heat. The method can preserve off-diagonal spin-exchange interactions.
\cref{fig4}(c) presents the experimental data of  $C_{mag}/T$ and the simulation results obtained for a 3 $\times$ 4 spin lattice with PBC. 
In comparison to the experimental data, the simulation results are consistent with the experimental data above 0.8 K. A noticeable deviation emerges below 0.8 K.
The deviation primarily stems from the limitations imposed by the finite lattice size. Longer-range spin correlations and quantum entanglement beyond the 3 $\times$ 4 spin lattice are neglected.
The off-diagonal spin-exchange interactions $J_{\pm}$ and $J_{z\pm}$ are determined to be $-$0.18 and 2.66 K, respectively.
With all parameters in the anisotropic spin-exchange interaction Hamiltonian now determined, we are poised for a comprehensive exploration of \ce{NaYbS2}'s ground-state magnetism.

\begin{figure*}[htb]
	\centering
	\includegraphics[scale=0.5]{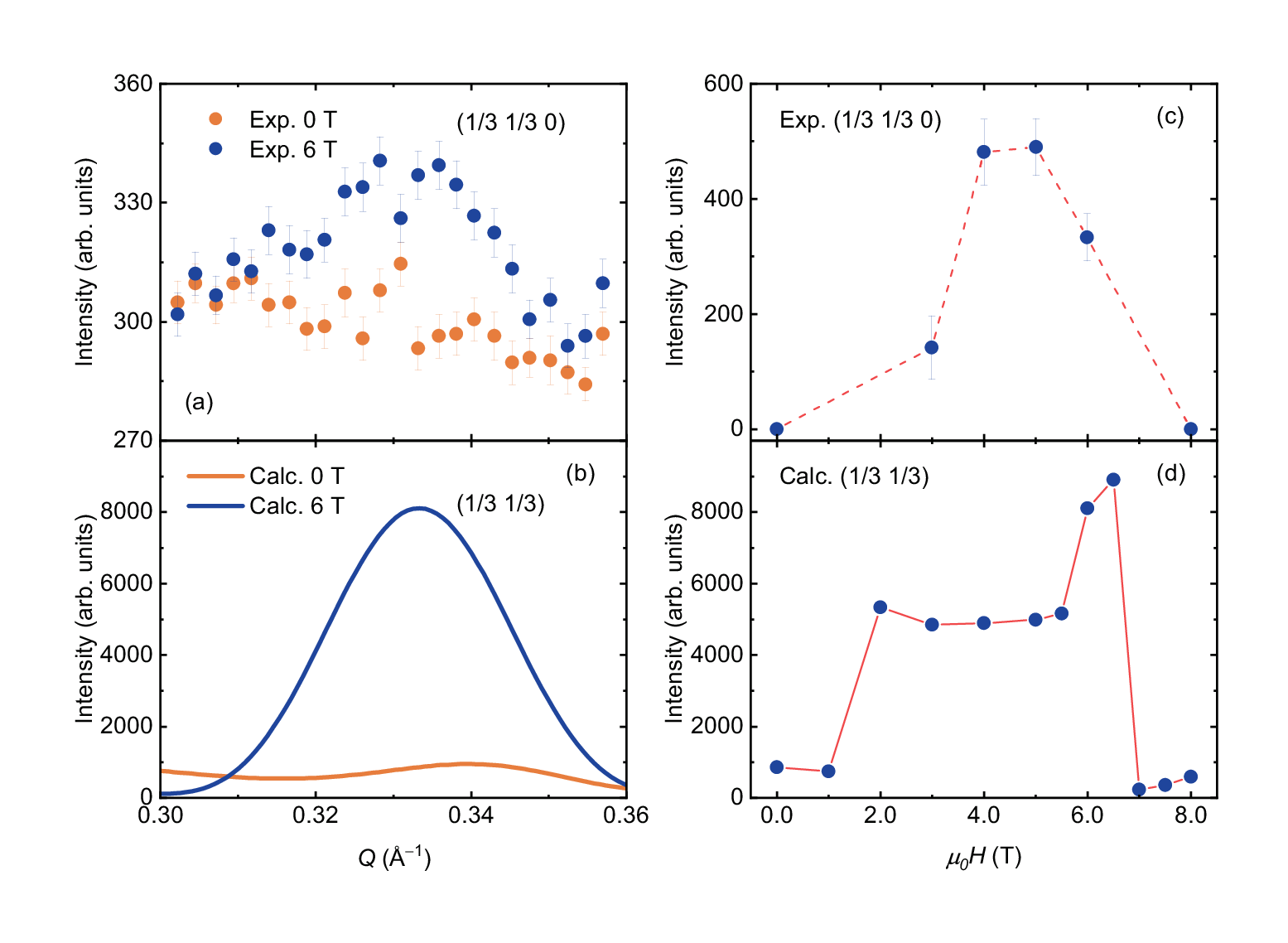}
	\caption{Neutron reflections\cite{Wu2022} and calculated results of static spin structure factor $S\left(\vec{q}\right)$ at $\vec{Q}$ = (1/3, 1/3, 0) for \ce{NaYbS2} ground state under different magnetic fields. (a) and (c) The comparison of the neutron reflections at (1/3, 1/3, 0) between 0 T and 6 T at 0.1 K and the magnetic field-dependence of (1/3, 1/3, 0), respectively. (b) and (d) The comparison of the $S\left(\vec{q}\right)$ at (1/3, 1/3) between 0 T and 6 T at ground state and the magnetic field-dependence of (1/3, 1/3), respectively.}
	\label{fig6}
\end{figure*}

\section{Ground state magnetism}
In this section, we employed the DMRG method to investigate the ground state magnetism of \ce{NaYbS2}.
The PBC was employed to get a accurate matrix product state (MPS). After benchmarking by selecting clusters of different sizes, we finally determined that there are no significant size effects for the $L_{x}$ $\times$ $L_{y}$ = 12 $\times$ 12 cluster.
In order to determine the location of \ce{NaYbS2} in the phase diagram, we first calculated the ground state phase diagram.
We set $J_{zz}$ as a dimensionless parameter of 1 and fixed the dimensionless parameters $J_{\pm}/J_{zz}$ = $-$4.05.
The parameter range for $J_{\pm\pm}/J_{zz}$ was chosen to be from $-$2 to 2, and the parameter range for $J_{z\pm}/J_{zz}$ is from 0 to 6.
Eventually, we obtained the two-dimensional ground state phase diagram.
\cref{fig5}(a) and \cref{fig5}(b) respectively display the ground state phase diagrams scaled by ground state energy and bipartite entanglement entropy.
The phase boundaries marked by white dashed lines in the ground state phase diagram are determined by bipartite entanglement entropy, the arrangement of ground state magnetic moments, and the static spin structure factor $\left\langle S_{i}S_{j} \right\rangle$.
Overall, the ground state phase diagram is in good agreement with existing results from the classical Monte Carlo simulations\cite{PhysRevB.94.035107,PhysRevB.95.165110}, exact diagonalization\cite{PhysRevB.95.165110,PhysRevB.103.205122}, and DMRG calculations\cite{PhysRevB.95.165110,PhysRevLett.120.207203,PhysRevX.9.021017}. 
We can divide the phase diagram into four regions: stripe-x, stripe-yz, AFM-$120^\circ$, and QSL phase.

\cref{fig5}(e), \cref{fig5}(f), and \cref{fig5}(i) show the periodic arrangement of static magnetic moments for AFM-$120^\circ$, stripe-x, and stripe-yz, respectively. 
For the stripe-x phase, the magnetic moments on each point are arranged in opposite directions along the edge of the triangle lattice.
The $S\left(\vec{q}\right)$ in reciprocal space shows sharp peak located at the high-symmetry point $M$ in the Brillouin zone (see \cref{fig5}(d)).
For the stripe-yz phase, similar to the stripe-x phase, the magnetic moments are arranged in opposite directions along the perpendicular direction of the triangular lattice. 
The peaks of $S\left(\vec{q}\right)$ are also located at the $M$ points (see \cref{fig5}(g)).
The bipartite entanglement entropy can provide us with more information. Compared to the QSL state and the AFM-$120^\circ$ phase, the entanglement entropy of the two stripe phases is close to zero.
This implies that the quantum fluctuations of these two stripe phases are weak.
The NN spin correlation $\left\langle S_{i}S_{j} \right\rangle$ also exhibits strong periodicity in the two stripe phases (see Supporting Information S10). Each spin lattice point forms a negative (blue), negative, and positive (red) correlation structure with the NN spin lattice points along the $a_{1}$, $a_{2}$, and $a_{3}$ directions, respectively.
AFM-$120^\circ$ is not a trivial quantum state, even though the magnetic moments form a long-range periodic arrangement in the ab-plane.
From the view of entanglement entropy, the bipartite entanglement of the AFM-$120^\circ$ phase is not low, indicating the presence of strong quantum fluctuations within this regime.
The NN spin correlation $\left\langle S_{i}S_{j} \right\rangle$ also exhibits some interesting properties (see Supporting Information S10).
The bonds with stronger correlations form a periodic hexagonal structure, while the spin site located at the center of the hexagonal structure has weak correlations with the surrounding NN spin sites.
This means that these central spins can generate confined spinon excitations with strong fluctuations. It is possible to realize a deconfined "liquid" phase by introducing next-nearest-neighbor or non-diagonal interactions.
More interestingly, AFM-$120^\circ$ phase can simultaneously break the translational symmetry of the triangular lattice and the U(1) spin rotation symmetry within a certain range of spin-exchange interactions\cite{PhysRevLett.95.237204,PhysRevLett.95.127207,Gao2022,Xiang2024}. 
The breaking of these two symmetries allows the spin system to keep strong quantum fluctuations, which is called a spin supersolid. Both spin wave excitations and continuum excitations can be observed in INS spectra.

Compared to ordered phases, we are more interested in the QSL phase located in the central region of the ground state diagram.
The phase has the strongest entanglement entropy in the phase diagram, which means that the spin system has strong quantum fluctuations due to anisotropic spin-exchange interactions.
The red star marks the position of \ce{NaYbS2} in the phase diagram which is located within the QSL phase region.
A series of related experiments, including neutron diffraction, INS, and $\mu$SR, have confirmed that the ground state of \ce{NaYbS2} is a quantum spin liquid state.
We attempted to understand the ground state of \ce{NaYbS2} from numerical calculations. We plotted the arrangement of magnetic moments in real space and the $S\left( \vec{q} \right)$ in reciprocal space calculated by DMRG, as shown in \cref{fig5}(j) and \cref{fig5}(h).
From the arrangement of magnetic moments, we found that the magnetic structure cannot be described by a single magnetic propagation vector, thus exhibiting the characteristics of spin disorder.
Such a magnetic configuration is more like a distorted AFM-$120^\circ$ phase, which also indicates that off-diagonal spin-exchange interactions play a key role in the formation of spin disorder.
More importantly, $S\left( \vec{q} \right)$ exhibits a broadened peak at the high-symmetry K point. One characteristic of Dirac-like spin liquids is the presence of weak peaks at the K points.\cite{PhysRevLett.120.207203}.
The NN spin correlation $\left\langle S_{i}S_{j} \right\rangle$ also shows randomness (see Supporting Information S10).
These calculation results can be compared with existing experimental results.
The zero-field neutron reflection of \ce{NaYbS2} reveals a weak magnetic Bragg peak near $\vec{Q}$ = $\left(1/3, 1/3, 0\right)$ (K point) at 0.1 K\cite{Wu2022} (see \cref{fig6}(a)).
At the same temperature, the neutron reflection under 6 T shows obvious magnetic Bragg peak at  $\vec{Q}$ = $\left(1/3, 1/3, 0\right)$, indicating that the magnetic field induces a magnetic ordered phase.
Based on the DMRG of 15 $\times$ 15 spin lattice with PBC, we calculated the $S\left( \vec{q} \right)$ of \ce{NaYbS2} under 0 T and 6 T, respectively (see \cref{fig6}(b)).
Similar to the experimental data, a weak peak and an obvious peak can be observed near $\vec{Q}$ = (1/3, 1/3) under 0 T and a 6 T, respectively.
The magnetic field-dependence neutron reflections at (1/3, 1/3, 0) are shown in \cref{fig6}(c)\cite{Wu2022}.
With the increasing magnetic field, the peak at (1/3, 1/3, 0) first increases and then decreases, which is consistent with the DMRG calculation results (see \cref{fig6}(d)).
Considering the neutron diffraction data was measured using polycrystalline samples of \ce{NaYbS2}, so the diffraction intensity might not be sufficient.
Besides, the limited data points of neutron diffraction under different magnetic fields may not fully reflect the trend of diffraction peaks with magnetic fields, especially the lack of data near quantum critical points.
To better illustrate the series of magnetic field-induced phase transitions in \ce{NaYbS2}, we also compare our calculated $S\left( \vec{q} \right)$ at $\vec{Q}$ = (1/3, 1/3) with the $\mathrm{d}M/\mathrm{d}H$ of \ce{NaYbS2}\cite{Wu2022}.
The $\mathrm{d}M/\mathrm{d}H$ data of \ce{NaYbS2} has a very significant change in the range of 6 to 6.5 T, which indicates that there is a field-induced phase transition in this magnetic field range. The peak of the $S\left( \vec{q} \right)$ at $\vec{Q}$ = (1/3, 1/3) also appears in the range of 6 to 6.5 T, which is consistent with the $\mathrm{d}M/\mathrm{d}H$ data. Therefore, our computational results essentially illustrate the evolution of the ground state magnetic structure of \ce{NaYbS2} with the magnetic field.
For INS, we can distinguish a cone feature at $|\vec{Q}|$ $\sim$ 1.24 \AA$^{-1}$ from powder data at 50 mK\cite{Wu2022}.
This core feature is also key evidence for the Dirac-like spin liquid.
From the perspective of spin dynamics, the ground state of \ce{NaYbS2} also confirms to the Dirac-like spin liquid\cite{PhysRevB.100.241116}. The zero-field $\mu$SR experimental data show that the spin relaxation rate maintains a finite value even at 0.1 K.
However, a weak magnetic field (about 0.1 T) can cause the spin relaxation rate to drop to near zero, significantly different from the LF-$\mu$SR spin relaxation rate of QSL with spinon Fermi surface state\cite{PhysRevX.11.021044,PhysRevB.106.085115}.
The former can open a spin gap at the Dirac point by applying a weak magnetic field, while the latter is robust against the influence of magnetic field.
The thermodynamic data of $C_{v}/T$ tends to 0 as the temperature decreases, rather than maintaining a finite value.
All of these numerical calculations and experimental results point to the ground state of \ce{NaYbS2} being a Dirac-like gapless QSL.

The consistency of experimental and computational results also demonstrates the validity of our chosen low-energy spin Hamiltonian model and the fitted exchange interactions.
In the future, we will conduct more in-depth INS measurements in single crystal samples of \ce{NaYbS2} to study the potential spinon excitations and topological properties.

\section{Summary}
In summary, we conducted a detailed investigation of the magnetism of \ce{NaYbS2} from finite temperatures to ground state.
Firstly, we conducted a Raman scattering on \ce{NaYbS2} using Raman spectroscopy techniques. 
Through polarized Raman scattering, we identified two phonon excitations with Raman activity, namely the $E_{g}$ and $A_{1g}$ modes.
More importantly, we clearly detected three CEF excitation energy levels of \ce{NaYbS2} at 2 K, which are at 207 $\mathrm{cm^{-1}}$ ($\sim$ 25.6 meV), 263 $\mathrm{cm^{-1}}$($\sim$ 32.6 meV), and 314 $\mathrm{cm^{-1}}$ ($\sim$ 38.9 meV).

Next, we performed relevant calculations on the contribution of CEF excitations to the magnetization. By fitting the $M/H-T$ data of \ce{NaYbS2} under a magnetic field of 0.1 T, we can extract the CEF parameters of \ce{NaYbS2}.
Based on the CEF parameters, we can obtain the excitation energy levels and wave functions by diagonalizing the CEF Hamiltonian. These energy levels are highly consistent with the results of Raman scattering, further verifying the reliability of these parameters we obtained from the fitting.
At the same time, we determined the characteristic temperature of \ce{NaYbS2} to be 40 K. Above the temperature, the CEF excitations play a key role in magnetism, and below the temperature, the spin-exchange interaction dominates.
This is consistent with the results obtained from the analysis of the ESR linewidths of \ce{NaYbS2} and \ce{NaYb_{0.1}Lu_{0.9}S2}.
We also studied the $g$-factors in the effective spin-1/2 model of \ce{NaYbS2}. Through ESR measurements of a dilute magnetic doped sample of \ce{NaYb_{0.1}Lu_{0.9}S2}, we determined the accurate $g$-factors in the ab-plane and along the c-axis to be 3.14 and 0.86, respectively.
Through the MF theory, we can well explain why there are differences in the $g$-factors of \ce{NaYbS2} and \ce{NaYb_{0.1}Lu_{0.9}S2} along the c-axis.

Before analyzing the ground state magnetism of \ce{NaYbS2}, we first obtained the spin-exchange interactions in the low-energy spin Hamiltonian through MF theory and FD method.
By fitting the $M/H-T$ data below the characteristic temperature, we can obtain the diagonal spin-exchange interactions $J_{\pm}$ and $J_{zz}$ to be 3.52 K and $-$0.87 K, respectively.
For the off-diagonal spin-exchange interactions, $J_{\pm\pm}$ and $J_{z\pm}$, we can obtain them to be $-$0.18 K and 2.66 K, respectively, by fitting the specific heat data in the low temperature range using the FD method.
We employed the DMRG method to investigate the ground state of \ce{NaYbS2}.
The calculation results directly point to the ground state of \ce{NaYbS2} being spin disordered. 
By comparing with the existing experimental results, we can further confirm that the ground state of \ce{NaYbS2} is a Dirac-like spin liquid.
This lays the foundation for us to further study the topological properties of the ground state magnetism of \ce{NaYbS2}.


\Acknowledgements{This work was supported by the National Key Research and Development Program of China (Grant No. 2022YFA1402704), the National Science Foundation of China (Grant No. 12274186),  the Strategic Priority Research Program of the Chinese Academy of Sciences (Grant No. XDB33010100), and the Synergetic Extreme Condition User Facility (SECUF). A portion of this work was performed on the Steady High Magnetic Field Facilities, High Magnetic Field Laboratory, CAS.}

\InterestConflict{The authors declare that they have no conflict of interest.}



\end{multicols}
\end{document}